\documentclass[12pt]{article}
\usepackage{epsfig,amsmath,euscript,graphicx}

\newcommand{\grw}{0.18\textwidth}
\newcommand{\gw}{0.15\textwidth}

\def\nslash{\rlap{\hspace{0.02cm}/}{n}}
\def\nbslash{\rlap{\hspace{0.02cm}/}{\bar n}}
\def\vslash{\rlap{\hspace{0.02cm}/}{v}}
\def\Dslash{\rlap{\hspace{0.07cm}/}{D}}

\def\H{{\EuScript H}}
\def\Q{{\EuScript Q}}
\def\X{{\EuScript X}}

\begin{document}

\begin{titlepage}

\begin{flushright}
CLNS~03/1843\\
SLAC--PUB--10166\\
{\tt hep-ph/0309227}
\end{flushright}

\vspace{0.7cm}
\begin{center}
\Large\bf 
External Operators and Anomalous Dimensions in Soft-Collinear 
Effective Theory
\end{center}

\vspace{0.8cm}
\begin{center}
{\sc T.~Becher$^{(a)}$, R.J.~Hill$^{(a)}$, B.O.~Lange$^{(b)}$, and 
M.~Neubert$^{(b)}$}\\
\vspace{0.7cm}
$^{(a)}${\sl Stanford Linear Accelerator Center, Stanford University\\
Stanford, CA 94309, U.S.A.} \\
\vspace{0.3cm}
$^{(b)}${\sl Newman Laboratory for Elementary-Particle Physics, 
Cornell University\\
Ithaca, NY 14853, U.S.A.}
\end{center}

\vspace{1.0cm}
\begin{abstract}
\vspace{0.2cm}\noindent
It has recently been argued that soft-collinear effective theory for 
processes involving both soft and collinear partons contains a new 
soft-collinear mode, which can communicate between the soft and collinear 
sectors of the theory. The formalism incorporating the corresponding 
fields into the effective Lagrangian is extended to include external 
current and four-quark operators relevant to weak interactions. An 
explicit calculation of the anomalous dimensions of these operators 
reveals that soft-collinear modes are needed for correctly describing 
the ultraviolet behavior of the effective theory.
\end{abstract}
\vfil

\end{titlepage}

\section{Introduction}

Soft-collinear effective theory (SCET) 
\cite{Bauer:2000yr,Bauer:2001yt,Chay:2002vy,Beneke:2002ph,Hill:2002vw} 
provides a systematic framework in which to discuss the factorization 
properties of exclusive $B$-decay amplitudes for processes in which the 
external hadronic states contain highly energetic, collinear partons 
inside light final-state mesons, and soft partons inside the initial $B$ 
meson. Power counting in SCET is based on an expansion parameter 
$\lambda\sim\Lambda/E$, where $E\gg\Lambda_{\rm QCD}$ is the large energy
carried by collinear particles (typically $E\sim m_b$ in $B$ decays) and 
$\Lambda\sim\Lambda_{\rm QCD}$ is of order the QCD scale. 

In a recent paper \cite{Becher:2003qh}, three of us have argued that the
intricate interplay between soft and collinear degrees of freedom makes 
it necessary to introduce modes with virtuality $E^2\lambda^3$, which 
have unsuppressed interactions with soft and collinear fields. In the 
strong-interaction sector of SCET, the leading-order couplings of these 
``soft-collinear'' fields to soft or collinear particles can be removed 
using field redefinitions, leaving residual interactions that are 
suppressed by at least two powers of $\lambda^{1/2}$. A puzzling aspect 
of this analysis was the finding that soft-collinear modes have 
virtualities that are parametrically below the QCD scale. We argued that 
this is to some extent a consequence of dimensional regularization and 
analyticity. What matters is not the virtuality but the fact that the 
plus and minus components of soft-collinear momenta are commensurate 
with certain components of collinear or soft momenta. Yet, one might 
worry whether the scaling laws derived for interactions of the 
soft-collinear fields might be invalidated by some non-perturbative 
effects, thereby upsetting the power counting of the effective theory. 

The goal of the present paper is to build up confidence in the new modes 
by showing explicitly that they are necessary to correctly reproduce the 
{\em ultraviolet\/} (UV) behavior of the effective theory, which has to 
match the scale dependence of short-distance coefficient functions 
derived in the matching of full-theory amplitudes onto matrix elements of 
SCET operators. Specifically, we compute the anomalous dimensions of the 
leading-order current operators containing a heavy and a collinear quark, 
a soft and a collinear quark, and four-quark operators obtained from 
combining these two currents. We find that without the inclusion of 
soft-collinear fields the results for the anomalous dimensions would be 
incorrect and violate fundamental principles of renormalization theory, 
such as the independence of renormalization-group (RG) functions of 
infrared (IR) regulators. 

The explicit examples we investigate exhibit two other important
features: first, in the presence of external operators such as
flavor-changing currents, the soft-collinear fields can in general no
longer be decoupled at leading order in $\lambda$ using field
redefinitions. Their effects must therefore be studied carefully in
applications of SCET to exclusive $B$ decays. This complicates proofs
of QCD factorization theorems. Secondly, only the sum of soft,
collinear and soft-collinear contributions to an amplitude is
physically meaningful. Through the particular scaling
$p_{sc}^2\sim\Lambda^3/E$ of soft-collinear momenta the amplitude
becomes sensitive to the large scale $E$. Part of this sensitivity has
a short-distance interpretation, as reflected in the anomalous
dimensions of SCET operators. However, in cases where the
soft-collinear modes cannot be decoupled, amplitudes may contain
additional dependence on the large scale that is of IR origin. In a
strongly coupled theory such as QCD this dependence cannot be
factorized using RG techniques.

\section{SCET fields and interactions at leading order}

The construction of an effective theory for collinear particles must 
account for the fact that different components of particle momenta and 
fields scale differently with the large scale $E$. To make this scaling 
explicit one introduces two light-like vectors $n^\mu$ and $\bar n^\mu$ 
satisfying $n^2=\bar n^2=0$ and $n\cdot\bar n=2$. Typically, 
$n^\mu=(1,0,0,1)$ is chosen to be the direction of an outgoing fast 
hadron (or a jet of hadrons), and $\bar n^\mu=(1,0,0,-1)$ points in the 
opposite direction. Any 4-vector can be decomposed as
\begin{equation}
   p^\mu = (n\cdot p)\,\frac{\bar n^\mu}{2}
    + (\bar n\cdot p)\,\frac{n^\mu}{2} + p_\perp^\mu
   \equiv p_+^\mu + p_-^\mu + p_\perp^\mu \,,
\end{equation}
where $p_\perp\cdot n=p_\perp\cdot\bar n=0$. This relation defines the 
light-like vectors $p_\pm^\mu$. The relevant SCET degrees of freedom 
describing the partons in the external hadronic states of exclusive $B$ 
decays are soft and collinear, where 
$p_s^\mu\sim E(\lambda,\lambda,\lambda)$ for soft momenta and 
$p_c^\mu\sim E(\lambda^2,1,\lambda)$ for collinear momenta. Here and 
below we indicate the scaling properties of the components 
$(n\cdot p,\bar n\cdot p,p_\perp)$. The corresponding effective-theory 
fields and their scaling relations are $h_v\sim\lambda^{3/2}$ (soft heavy 
quark), $q_s\sim\lambda^{3/2}$ (soft light quark), 
$A_s^\mu\sim(\lambda,\lambda,\lambda)$ (soft gluon), and 
$\xi\sim\lambda$ (collinear quark), $A_c^\mu\sim(\lambda^2,1,\lambda)$ 
(collinear gluon). At leading order in power counting the effective 
strong-interaction Lagrangian splits up into separate Lagrangians for the 
soft and collinear fields. However, as mentioned above, the effective 
theory also contains soft-collinear quark and gluon fields, 
$\theta\sim\lambda^2$ and 
$A_{sc}^\mu\sim(\lambda^2,\lambda,\lambda^{3/2})$, which have 
leading-order couplings to both soft and collinear fields. The formalism
incorporating these fields has been developed in \cite{Becher:2003qh},
borrowing methods developed by Beneke and Feldmann \cite{Beneke:2002ni}. 
It will be briefly reviewed here. 

In interactions with other fields, the soft-collinear fields (but not the 
soft and collinear fields) are multipole expanded as
\begin{equation}
\begin{aligned}
   \phi_{sc}(x) &= \phi_{sc}(x_-) + x_\perp\cdot\partial_\perp\,
    \phi_{sc}(x_-) + \dots \quad && \mbox{in collinear interactions}
    \,, \\
   \phi_{sc}(x) &= \phi_{sc}(x_+) + x_\perp\cdot\partial_\perp\,
   \phi_{sc}(x_+) + \dots \quad && \mbox{in soft interactions} \,.
\end{aligned}
\end{equation}
The first correction terms are of $O(\lambda^{1/2})$, and the omitted 
terms are of $O(\lambda)$ and higher. Soft-collinear fields can couple to 
soft or collinear fields without altering their scaling properties. This 
motivates the treatment of the soft-collinear gluon field as a background 
field. However, in order to preserve the scaling properties of the fields 
under gauge transformations one must expand the transformation laws in
$\lambda$. This leads to the following set of ``homogeneous'' gauge 
transformations for the quark fields:
\begin{equation}\label{SCtrafo}
\begin{aligned}
   \mbox{soft:} \quad &
    q_s(x)\to U_s(x)\,q_s(x) \,, &&
    \mbox{collinear and soft-collinear fields invariant} \\
   \mbox{collinear:} \quad &
    \xi(x)\to U_c(x)\,\xi(x) \,, &&
    \mbox{soft and soft-collinear fields invariant} \\
   \mbox{soft-collinear:} \quad & 
    q_s(x)\to U_{sc}(x_+)\,q_s(x) \,, ~\,&&
    \xi(x)\to U_{sc}(x_-)\,\xi(x) \,, \quad
    q_{sc}(x)\to U_{sc}(x)\,q_{sc}(x)
\end{aligned}
\end{equation}
The transformation laws for gluons \cite{Becher:2003qh} are more
complicated and are not needed for the present work.

The effective Lagrangian of SCET can be split up as
\begin{equation}\label{Lscet}
   {\cal L}_{\rm SCET}
   = {\cal L}_s + {\cal L}_c + {\cal L}_{sc}
   + {\cal L}_{\rm int}^{(0)} + \dots \,,
\end{equation}
where the dots represent power-suppressed interaction terms. The 
integration measure $d^4x$ in the action 
$S_{\rm SCET}=\int d^4x\,{\cal L}_{\rm SCET}$ scales like $\lambda^{-4}$
for all terms except the soft-collinear Lagrangian ${\cal L}_{sc}$, for
which it scales like $\lambda^{-6}$. The first three terms above 
correspond to the Lagrangians of soft particles (including heavy quarks), 
collinear particles, and soft-collinear particles. They are given by
\begin{equation}
\begin{aligned}
   {\cal L}_s &= \bar q_s\,i\Dslash_s\,q_s + \bar h\,iv\cdot D_s\,h
    + {\cal L}_s^{\rm glue} \,, \\[0.1cm]
   {\cal L}_c
   &= \bar\xi\,\frac{\nbslash}{2}\,in\cdot D_c\,\xi
    - \bar\xi\,i\Dslash_{c\perp}\,\frac{\nbslash}{2}\,
    \frac{1}{i\bar n\cdot D_c}\,i\Dslash_{c\perp}\,\xi 
    + {\cal L}_c^{\rm glue} \,, \\[-0.1cm]
   {\cal L}_{sc}
   &= \bar\theta\,\frac{\nbslash}{2}\,in\cdot D_{sc}\,\theta
    - \bar\theta\,i\Dslash_{sc\perp}\,\frac{\nbslash}{2}\,
    \frac{1}{i\bar n\cdot D_{sc}}\,i\Dslash_{sc\perp}\,\theta
    + {\cal L}_{sc}^{\rm glue} \,,
\end{aligned}
\end{equation}
where $iD_s^\mu\equiv i\partial^\mu+g A_s^\mu$ etc., and $v$ is the 
velocity of the hadron containing the heavy quark. Collinear, 
soft-collinear, and heavy-quark fields in the effective theory are 
described by 2-component spinors subject to the constraints 
$\nslash\,\xi=0$, $\nslash\,\theta=0$, and $\vslash\,h=h$. The gluon 
Lagrangians in the three sectors retain the same form as in full QCD, 
but with the gluon fields restricted to the corresponding subspaces of 
their soft, collinear, or soft-collinear Fourier modes. The leading-order 
interactions between soft-collinear fields and soft or collinear fields 
are given by
\begin{eqnarray}\label{SCints}
   {\cal L}_{\rm int}^{(0)}(x)
   &=& \bar q_s(x)\,\frac{\nslash}{2}\,g\bar n\cdot A_{sc}(x_+)\,q_s(x)
    + \bar h(x)\,\frac{n\cdot v}{2}\,g\bar n\cdot A_{sc}(x_+)\,h(x)
    \nonumber\\
   &+& \bar\xi(x)\,\frac{\nbslash}{2}\,g n\cdot A_{sc}(x_-)\,
    \xi(x) + \mbox{pure glue terms} \,.
\end{eqnarray}
Momentum conservation implies that soft-collinear fields can only
couple to either soft or collinear modes, but not both. More than one
soft or collinear particle must be involved in such interactions. The
gluon self-couplings can be derived by substituting $A_s^\mu\to
A_s^\mu+\frac12 n^\mu\,\bar n\cdot A_{sc}(x_+)$ for the gluon field in
the soft Yang--Mills Lagrangian and $A_c^\mu\to A_c^\mu+\frac12 \bar
n^\mu\,n\cdot A_{sc}(x_-)$ for the gluon field in the collinear
Yang--Mills Lagrangian, and isolating terms containing the
soft-collinear field. The precise form of these interactions will not
be relevant to our discussion. Finally, let us note that none of the
terms in the SCET Lagrangian (\ref{Lscet}) is renormalized beyond the
usual renormalization of the strong coupling and the fields.

From (\ref{SCints}) one can readily read off the Feynman rules for the 
couplings of soft-collinear gluons to soft or collinear quarks. The 
multipole expansion of the soft-collinear fields implies that momentum 
is {\em not\/} conserved at these vertices. When a soft (light or 
heavy) quark with momentum $p_s$ absorbs a soft-collinear gluon with 
momentum $k$, the outgoing soft quark carries momentum $p_s+k_-$. 
Likewise, when a collinear quark with momentum $p_c$ absorbs a 
soft-collinear gluon with momentum $k$, the outgoing collinear quark 
carries momentum $p_c+k_+$.

In order to match the quark and gluon fields of the full theory onto 
SCET fields obeying the homogeneous gauge transformations one first 
adopts specific gauges in the soft and collinear sectors, namely soft 
light-cone gauge $n\cdot A_s=0$ (SLCG) and collinear light-cone gauge 
$\bar n\cdot A_c=0$ (CLCG). At leading order in $\lambda$, one then 
introduces the corresponding SCET fields via the substitutions 
\cite{Beneke:2002ni}
\begin{equation}\label{newfields}
   \psi_s \big|_{\rm SLCG} \to R_s\,S_s^\dagger\,q_s \,, \qquad
   b \big|_{\rm SLCG} \to R_s\,S_s^\dagger\,h \,, \qquad
   \psi_c \big|_{\rm CLCG} \to R_c\,W_c^\dagger\,\xi \,.
\end{equation}
The corresponding replacements for gluon fields can be found in 
\cite{Becher:2003qh}. The quantities
\begin{equation}\label{WSdef}
\begin{aligned}
   S_s(x) &= \mbox{P}\exp\left( ig\int_{-\infty}^0 dt\,
    n\cdot A_s(x+tn) \right) , \\
   W_c(x) &= \mbox{P}\exp\left( ig\int_{-\infty}^0 dt\,
    \bar n\cdot A_c(x+t\bar n) \right) 
\end{aligned} 
\end{equation}
are the familiar SCET Wilson lines in the soft and collinear sectors
\cite{Bauer:2001yt,Beneke:2002ph}, which effectively put the SCET
fields into light-cone gauge. The objects $R_s$ and $R_c$ are short
gauge strings of soft-collinear fields from $x_+$ to $x$ (for $R_s$)
and $x_-$ to $x$ (for $R_c$). They differ from 1 by terms of order
$\lambda^{1/2}$ and so must be Taylor expanded. Note that $S_s$
transforms as $S_s(x)\to U_s(x)\,S_s(x)$ and $S_s(x)\to
U_{sc}(x_+)\,S_s(x)\,U_{sc}^\dagger(x_+)$ under soft and
soft-collinear gauge transformations and is invariant under collinear
gauge transformations. Likewise, $W_c$ transforms as $W_c(x)\to
U_c(x)\,W_c(x)$ and $W_c(x)\to
U_{sc}(x_-)\,W_c(x)\,U_{sc}^\dagger(x_-)$ under collinear and
soft-collinear gauge transformations and is invariant under soft gauge
transformations. The short strings only transform under soft-collinear
gauge transformations, in such a way that $R_s(x)\to
U_{sc}(x)\,R_s(x)\,U_{sc}^\dagger(x_+)$ and $R_c(x)\to
U_{sc}(x)\,R_c(x)\,U_{sc}^\dagger(x_-)$. It follows that the
expressions on the right-hand side of (\ref{newfields}) are invariant
under soft and collinear gauge transformations and transform as
ordinary QCD quark fields under soft-collinear gauge transformations.

\section{Soft-collinear current operators}

Flavor-changing currents and four-quark operators containing soft and 
collinear fields play an important role in many applications of SCET to 
exclusive $B$ decays. The simplest example is that of a current 
$\bar\psi_c\,\Gamma\,b$ transforming a heavy quark into a collinear one,
where $\Gamma$ denotes an arbitrary Dirac structure. This current has 
been studied in detail in \cite{Bauer:2000yr,Beneke:2002ph} in another
version of SCET, which contains only hard-collinear and soft 
fields.\footnote{This theory is sometimes called SCET$_{\rm I}$, and 
its degrees of freedom are often called collinear and ultra-soft. The
effective theory considered in the present paper is also called 
SCET$_{\rm II}$.} Based on the discussion of the previous section, it 
follows that at tree level (and at leading power in $\lambda$) the QCD 
current is matched onto the following gauge-invariant object in SCET:
\begin{eqnarray}\label{hc}
   \bar\psi_c(x)\,\Gamma\,b(x)
   &\to& e^{-im_b v\cdot x}\,\big[ \bar\xi\,W_c\,R_c^\dagger \big](x)\,
    \Gamma\,\big[ R_s\,S_s^\dagger\,h \big](x) \nonumber\\
   &=& e^{-im_b v\cdot x}\,\big[ \bar\xi\,W_c \big](x_+ + x_\perp)\,
    \Gamma\,\big[ S_s^\dagger\,h \big](x_- + x_\perp) + O(\lambda) \,,
\end{eqnarray}
where the phase factor arises from the definition of the field $h$ in
HQET \cite{Neubert:1993mb}. Note that the expression in the first line 
is not homogeneous in $\lambda$. In interactions of soft and collinear 
fields, the soft fields must be multipole expanded about $x_+=0$, while 
the collinear fields must be multipole expanded about $x_-=0$. Also, as
mentioned above, the quantities $R_s$ and $R_c$ must be expanded and 
equal 1 to first order. This leads to the result shown in the second 
line. The terms of $O(\lambda^{1/2})$ in the expansions of $R_s$ and 
$R_c$ cancel each other. The leading-order SCET current in the final 
expression is gauge invariant even without the $R_s$ and $R_c$ factors, 
since according to (\ref{SCtrafo}) soft fields at $x_+=0$ and collinear
fields at $x_-=0$ both transform with $U_{sc}(0)$ under soft-collinear 
gauge transformations. An analogous matching relation can be written for
a soft-collinear current containing a light soft quark, 
\begin{equation}\label{sc}
   \bar\psi_c(x)\,\Gamma\,\psi_s(x)
   \to \big[ \bar\xi\,W_c \big](x_+ + x_\perp)\,\Gamma\,
   \big[ S_s^\dagger\,q_s\big](x_- + x_\perp) + O(\lambda) \,.
\end{equation}
Depending on the Dirac structure $\Gamma$, another operator containing 
an additional perpendicular collinear gluon field can appear in this case 
(even at tree level) \cite{Hill:2002vw}. We will not discuss such 
operators in the present paper.

When radiative corrections are taken into account, the currents in 
(\ref{hc}) and (\ref{sc}) mix with analogous operators at different 
positions on the light cone, and for the case of the heavy-collinear 
current different Dirac structures can be induced by hard gluon exchange. 
The correct matching relations are (setting $x=0$ for simplicity) 
\cite{Beneke:2002ph,Hill:2002vw}
\begin{equation}\label{fullmatch}
\begin{aligned}
   \bar\psi_c(0)\,\Gamma\,b(0)
   &\to \sum_i \int ds\,\widetilde C_i(s,\mu)\,
    \big[ \bar\xi\,W_c \big](s\bar n)\,\Gamma_i\,
    \big[ S_s^\dagger\,h \big](0) + O(\lambda) \,, \\
   \bar\psi_c(0)\,\Gamma\,\psi_s(0)
   &\to \int ds dt\,\widetilde D(s,t,\mu)\,
    \big[ \bar\xi\,W_c \big](s\bar n)\,\Gamma\,
    \big[ S_s^\dagger\,q_s \big](tn) + O(\lambda) \,.
\end{aligned}
\end{equation}
Translational invariance can be used to rewrite these relations in the 
local form
\begin{equation}
\begin{aligned}
   \big[ \bar\psi_c\,\Gamma\,b\, \big](0)
   &\to \sum_i C_i(v\cdot P_-^c,\mu)\,
    \big[ \bar\xi\,W_c\,\Gamma_i\,S_s^\dagger\,h\big](0)
    + O(\lambda) \,, \\
   \big[ \bar\psi_c\,\Gamma\,\psi_s \big](0)
   &\to D(P_+^s\!\cdot P_-^c,\mu)\,
    \big[ \bar\xi\,W_c\,\Gamma\,S_s^\dagger\,q_s \big](0)
    + O(\lambda) \,,
\end{aligned}
\end{equation}
where
\begin{equation}
\begin{aligned}
   C_i(v\cdot P_-^c,\mu)
   &= \int ds\,\widetilde C_i(s,\mu)\,e^{is\bar n\cdot P^c} \,, \\
   D(P_+^s\!\cdot P_-^c,\mu)
   &= \int ds dt\,\widetilde D(s,t,\mu)\,
    e^{i(s\bar n\cdot P^c - tn\cdot P^s)}
\end{aligned}
\end{equation}
are the Fourier transforms of the position-space Wilson coefficients. 
These are operator-valued coefficient functions, which depend on the 
momentum operators $P^c=P_{\rm out}^c-P_{\rm in}^c$ and 
$P^s=P_{\rm in}^s-P_{\rm out}^s$ acting on collinear and soft states.
Invariance of the results under simultaneous rescalings $n\to n/\alpha$ 
and $\bar n\to\alpha\bar n$ of the light-cone basis vectors dictates that 
the momentum-space Wilson coefficients can only depend on the scalar 
products $2v\cdot P_-^c=(v\cdot n)\,(P^c\cdot\bar n)$ and 
$2P_+^s\!\cdot P_-^c=(P^s\cdot n)\,(P^c\cdot\bar n)$.

The momentum-space coefficient functions are renormalized 
multiplicatively and obey RG equations of the Sudakov type 
\cite{Bauer:2000yr,Bosch:2003fc},
\begin{equation}
\begin{aligned}
   \frac{d}{d\ln\mu}\,C_i(v\cdot p_{c-},\mu)
   &= \gamma_{\xi h}(v\cdot p_{c-},\mu)\,C_i(v\cdot p_{c-},\mu) \,, \\
   \frac{d}{d\ln\mu}\,D(p_{s+}\!\cdot p_{c-},\mu)
   &= \gamma_{\xi q}(p_{s+}\!\cdot p_{c-},\mu)\,
    D(p_{s+}\!\cdot p_{c-},\mu) \,, 
\end{aligned}
\end{equation}
where the anomalous dimensions take the form
\begin{equation}\label{andims}
\begin{aligned}
   \gamma_{\xi h}(v\cdot p_{c-},\mu)
   &= -\frac12\,\Gamma_{\rm cusp}[\alpha_s(\mu)]\,
    \ln\frac{\mu^2}{(2v\cdot p_{c-})^2}
    + \Gamma_{\xi h}[\alpha_s(\mu)] \,, \\
   \gamma_{\xi q}(p_{s+}\!\cdot p_{c-},\mu)
   &= -\Gamma_{\rm cusp}[\alpha_s(\mu)]\,
    \ln\frac{\mu^2}{2p_{s+}\!\cdot p_{c-}}
    + \Gamma_{\xi q}[\alpha_s(\mu)] \,.
\end{aligned}
\end{equation}
The coefficients of the logarithmic terms are determined in terms of 
the universal cusp anomalous dimension 
$\Gamma_{\rm cusp}=C_F\,\alpha_s/\pi+O(\alpha_s^2)$, which plays a 
central role in the renormalization of Wilson lines with light-like 
segments \cite{Korchemsky:wg}. In Section \ref{sec:decoupling} we will 
show why the cusp anomalous dimension enters the above equation with a 
negative sign, and why an additional factor of $1/2$ appears in the 
anomalous dimension of the heavy-collinear current.

 The one-loop expressions for the non-logarithmic
terms in the anomalous dimensions can be deduced from the explicit
results for the Wilson coefficients derived in
\cite{Bauer:2000yr,Hill:2002vw}.\footnote{The Wilson coefficients
$C_i(v\cdot p_{c-},\mu)$ were computed in \cite{Bauer:2000yr} using 
the effective theory SCET$_{\rm I}$, in which the the scaling of the
collinear quark field relative to the heavy-quark field is different
from that in the theory SCET$_{\rm II}$ considered here. The Wilson
coefficients are the same in the two theories because they are
independent of $p_c^2$, and so the scaling of the collinear fields
does not matter.}
They are
\begin{equation}\label{Gammas}
   \Gamma_{\xi h}(\alpha_s) = -\frac54\,\frac{C_F\alpha_s}{\pi}
    + O(\alpha_s^2) \,, \qquad
   \Gamma_{\xi q}(\alpha_s) = -\frac32\,\frac{C_F\alpha_s}{\pi}
    + O(\alpha_s^2) \,.
\end{equation}

It may seem surprising that after hard and hard-collinear scales have
been integrated out the operators of the low-energy theory still know
about the large scales $v\cdot p_{c-}\sim E$ and $p_{s+}\cdot
p_{c-}\sim E\Lambda$, as is evident from the appearance of the
logarithms in (\ref{andims}). The reason is that in interactions
involving both soft and collinear particles there is a large Lorentz
boost $\gamma\sim p_s\cdot p_c/\sqrt{p_s^2\,p_c^2}\sim E/\Lambda$
connecting the rest frames of soft and collinear hadrons, which is
fixed by external kinematics and enters the effective theory as a
parameter. This is similar to applications of heavy-quark effective
theory to $b\to c$ transitions, where the fields depend on the
external velocities of the hadrons containing the heavy quarks, and
$\gamma=v_b\cdot v_c=O(1)$ is an external parameter that appears in
matrix elements and anomalous dimensions of velocity-changing current
operators \cite{Neubert:1993mb,Falk:1990yz}. 

\begin{figure}
\begin{center}
\begin{tabular}{cccccc}
\includegraphics[width=\grw]{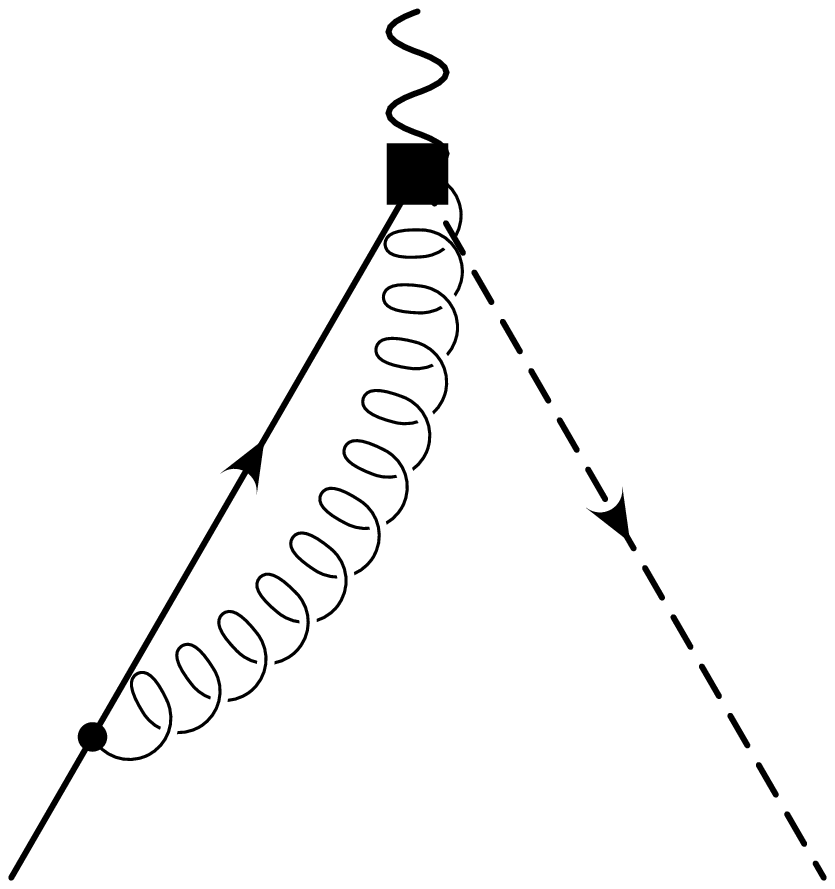} && 
\includegraphics[width=\grw]{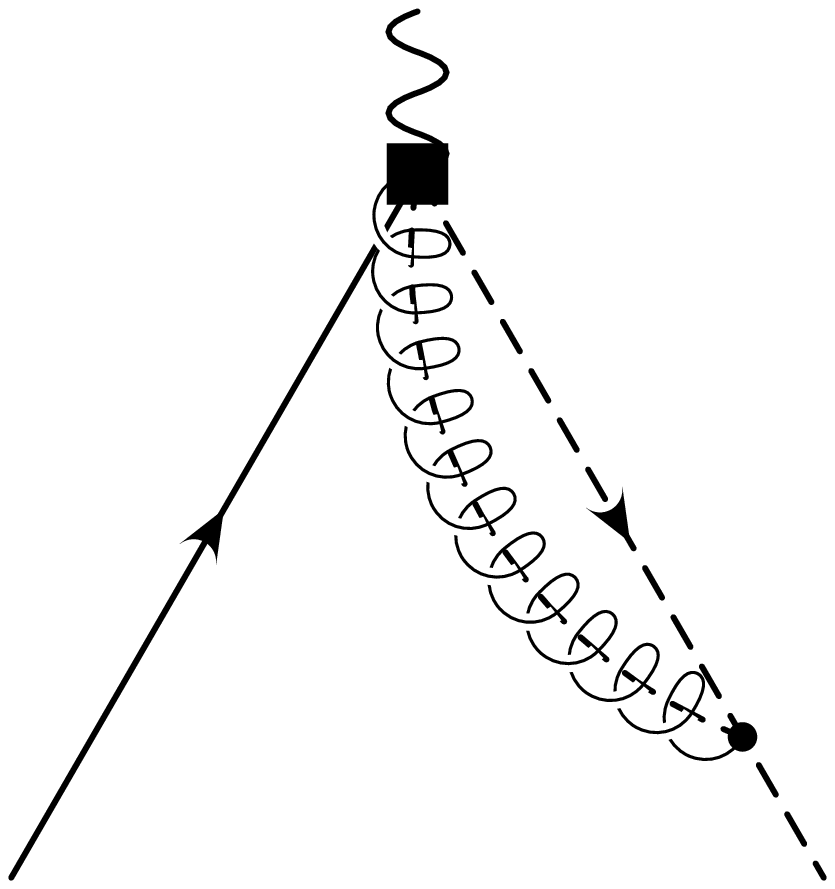} &&
\includegraphics[width=\grw]{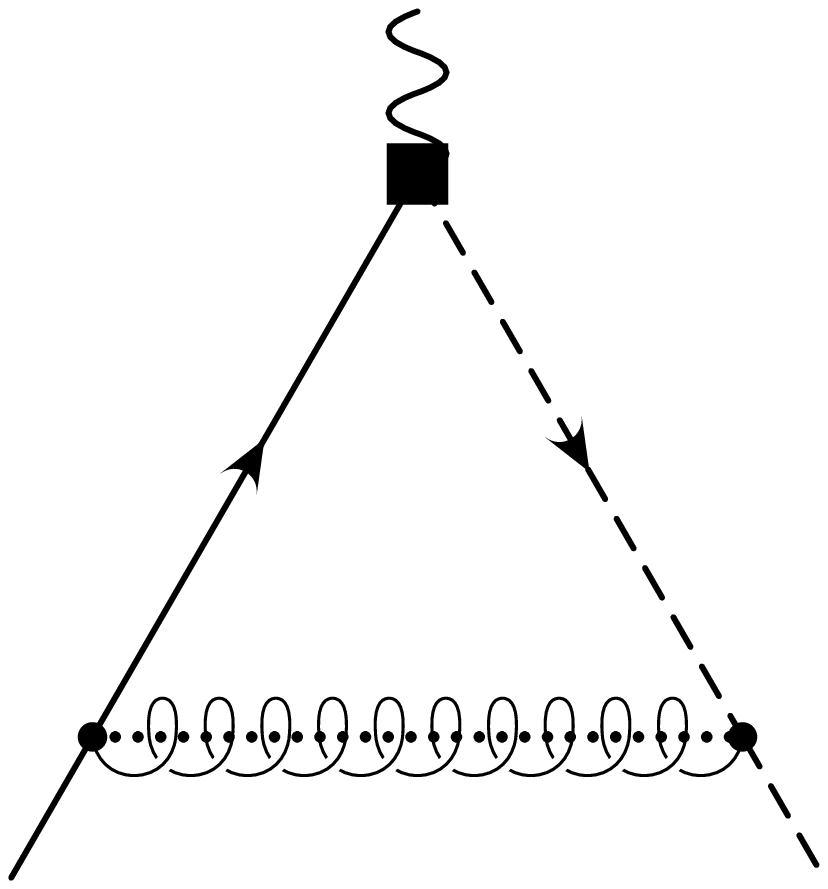}
\end{tabular}
\end{center}
\vspace{-0.2cm}
\centerline{\parbox{14cm}{\caption{\label{fig:SCETgraphs}
SCET graphs contributing to the anomalous dimension of a soft-collinear 
current. Full lines denote soft fields, dashed lines collinear fields, 
and dotted lines soft-collinear fields.}}}
\end{figure}

We will now explain how the results for the anomalous dimensions can be 
obtained from a calculation of UV poles of SCET loop diagrams. The 
relevant diagrams needed at one-loop order are shown in 
Figure~\ref{fig:SCETgraphs}. They must be supplemented by wave-function 
renormalization of the quark fields. The gluons connected to the current 
are part of the Wilson lines $W_c$ and $S_s$. We regularize IR 
singularities by keeping the external lines off-shell. The results for 
the sum of all UV poles must be independent of the IR regulators. For the 
heavy-collinear current the pole terms obtained from the three diagrams 
are (here and below we omit the $-i0$ in the arguments of logarithms)
\begin{eqnarray}\label{eq:hc}
   \left( \frac{1}{\epsilon^2}
   - \frac{2}{\epsilon}\,\ln\frac{-2v\cdot p_s}{\mu}
   + \frac{1}{\epsilon} \right)
   &+& \left( \frac{2}{\epsilon^2}
   - \frac{2}{\epsilon}\,\ln\frac{-p_c^2}{\mu^2}
   + \frac{3}{2\epsilon} \right)
   + \left( - \frac{2}{\epsilon^2} + \frac{2}{\epsilon}\,
   \ln\frac{(-2v\cdot p_s)(-p_c^2)}{2v\cdot p_{c-}\,\mu^2} \right)
   \nonumber\\
   &=& \frac{1}{\epsilon^2}
   + \frac{2}{\epsilon}\,\ln\frac{\mu}{2v\cdot p_{c-}}
   + \frac{5}{2\epsilon} \,,
\end{eqnarray}
while for the current containing a light soft quark we obtain
\begin{eqnarray}\label{eq:sc}
   \left( \frac{2}{\epsilon^2}
   - \frac{2}{\epsilon}\,\ln\frac{-p_s^2}{\mu^2}
   + \frac{3}{2\epsilon} \right)
   &+& \left( \frac{2}{\epsilon^2}
   - \frac{2}{\epsilon}\,\ln\frac{-p_c^2}{\mu^2}
   + \frac{3}{2\epsilon} \right)
   + \left( - \frac{2}{\epsilon^2} + \frac{2}{\epsilon}\,
   \ln\frac{(-p_s^2)(-p_c^2)}{2p_{s+}\!\cdot p_{c-}\,\mu^2} \right)
   \nonumber\\
   &=& \frac{2}{\epsilon^2}
   + \frac{2}{\epsilon}\,\ln\frac{\mu^2}{2p_{s+}\!\cdot p_{c-}}
   + \frac{3}{\epsilon} \,.
\end{eqnarray}
We quote the contributions to the operator renormalization constants
$Z^{-1}$ in units of $C_F\alpha_s/4\pi$ (in the $\overline{\rm MS}$
subtraction scheme with $D=4-2\epsilon$). The three parentheses in the
first line of the above equations correspond to the soft, collinear,
and soft-collinear contributions, where the first two terms include
the corresponding contributions from wave-function
renormalization. Note that the $1/\epsilon$ poles of the soft and
collinear graphs depend on the IR regulators, but that this dependence
is precisely canceled by the soft-collinear contribution. By
construction, the sum of the soft, collinear, and soft-collinear
contributions is IR finite and only contains UV poles, whose
coefficients depend on the ratios $v\cdot p_{c-}/\mu$ and
$p_{s+}\!\cdot p_{c-}/\mu^2$.  This follows since IR divergences in
both the full and the effective theory (which are equivalent at low
energy) are regularized by the off-shellness of the external quark
lines. The one-loop contributions to the anomalous dimensions
$\gamma_{\xi h}$ and $\gamma_{\xi q}$ are given by $-C_F\alpha_s/2\pi$
times the coefficients of the $1/\epsilon$ poles in the above
expressions. They are in agreement with the results (\ref{andims}) and
(\ref{Gammas}) obtained from the scale dependence of Wilson
coefficients.

The calculations presented above make it evident that there is an
intricate interplay between the soft, collinear, and soft-collinear
diagrams. In dimensional regularization the dependence of the
anomalous dimensions on the hard or hard-collinear scale enters
through the loop integral involving the soft-collinear exchange and
thus seems to be related to very small momentum scales. However, care
must be taken when assigning physical significance to the scales
associated with individual diagrams in SCET, because in the soft and
collinear diagrams a cancellation of IR and UV divergences takes
place. The logarithms appearing in their divergent parts should be
interpreted as [cf.~(\ref{eq:sc})]
\begin{equation}
   \ln\frac{-p_s^2}{\mu^2} + \ln\frac{-p_c^2}{\mu^2}
   = \ln\frac{Q^2}{\mu^2} + \ln\frac{m_{sc}^2}{\mu^2} \,, 
   \qquad \mbox{with} \quad
   m_{sc}^2 = \frac{(-p_s^2)\,(-p_c^2)}{2 p_{s+}\!\cdot p_{c-}} \,,
\end{equation}
and thus arise from a cancellation of physics at the hard scale 
$Q^2=2p_{s+}\!\cdot p_{c-}$ and at the soft-collinear scale $m_{sc}^2$.
In the sum of all graphs, the soft-collinear contribution precisely 
cancels the IR piece of the soft and collinear parts, see (\ref{eq:sc}).
This interpretation is consistent with the fact that the anomalous
dimensions measure the change of operator matrix elements under
infinitesimal variations of the UV cutoff $\mu$. They are therefore
insensitive to the physics at low scales by construction, and
the large logarithms in (\ref{andims}) are really of short-distance
nature. 

On the other hand, in the sum of the finite terms of the
diagrams in Figure~\ref{fig:SCETgraphs} (corresponding to SCET matrix
elements) logarithms of the soft-collinear scale remain, which do not
have an interpretation in terms of RG logarithms. In a weakly coupled
theory such as QED, the large logarithms
$\ln(\mu^2/m_{sc}^2)\sim\ln(\mu^2/(\Lambda^3/E))$ can be summed by
matching SCET onto another effective theory, in which soft and
collinear fields are integrated out and only soft-collinear fields
remain as dynamical degrees of freedom, and by solving RG equations in
this final theory. This is analogous to the evolution equations for
the off-shell Sudakov form factor discussed in
\cite{Korchemsky:1988hd,Kuhn:1999nn}, where large logarithms arise
from two-stage evolution between the scales $Q^2$ to $M^2$ and $M^2$
to $M^4/Q^2$. As argued in \cite{Becher:2003qh}, the case of the
current $\bar\psi_c\,\Gamma\,\psi_s$ can be mapped onto the Sudakov
problem, such that $Q^2$ corresponds to the hard-collinear scale,
$M^2$ corresponds to the QCD scale, and $M^4/Q^2$ corresponds to the
soft-collinear scale. Since QCD is strongly coupled for scales of
order $\Lambda$ and below the second stage of running cannot be 
performed perturbatively, i.e., the low-energy hadronic matrix 
elements in SCET may contain a dependence on the small ratio 
$\Lambda/E$ that cannot be factorized into a short-distance 
coefficient.

The observation that the soft-collinear contribution supplies a
logarithm of a short-distance scale solves the following puzzle about
soft-collinear current operators in SCET (which has confused some of
the present authors for a considerable amount of time): We know from 
the explicit expressions for the Wilson coefficients of the currents 
that their anomalous dimensions must depend on a scalar product of the 
collinear momentum with a momentum characterizing the soft quark (i.e., 
$v\cdot p_{c-}$ or $p_{s+}\!\cdot p_{c-}$). However, the SCET Feynman 
rules imply that the first graph in Figure \ref{fig:SCETgraphs} can only 
be a function of the soft momentum (i.e., $v\cdot p_s$ or $p_s^2$), 
while the second one can only depend on the collinear momentum (i.e.,
$p_c^2$), as is in fact confirmed by our explicit calculation. The
apparent ``factorization'' of soft and collinear degrees of freedom in
SCET (in the absence of soft-collinear fields) would thus lead to the
conclusion that the anomalous dimensions of the currents are 
independent of the products $v\cdot p_{c-}$ or $p_{s+}\cdot p_{c-}$,
in contradiction with the results for the Wilson coefficients. It is 
possible to obtain the correct result
for the anomalous dimensions by using a regularization scheme that
breaks this factorization property; however, unavoidably this means
that the regulator cannot preserve the symmetries of the Lagrangian of
the effective theory. (For instance, it is possible to suppress the
soft-collinear contribution in one-loop graphs by putting the external
lines on shell and using a different IR regulator such as a gluon
mass, in which case the last diagram in Figure~\ref{fig:SCETgraphs}
vanishes in dimensional regularization, whereas the first two diagrams
give expressions that cannot be regularized dimensionally. To give
meaning to these expressions one may introduce additional analytic
regulators, which break factorization and gauge invariance.)  The
formalism employing soft-collinear fields provides an elegant solution
to this problem by making the non-factorization of the soft and
collinear sectors of SCET explicit at the level of the Lagrangian,
avoiding any subtleties related to regularization.

\section{Four-quark operators}

In many cases relevant to $B$ physics, amplitudes calculated in SCET 
receive contributions from hadronic matrix elements of four-quark 
operators, which can be expressed in terms of the leading-order 
light-cone distribution amplitudes (LCDAs) of a light final-state meson 
and of the initial state $B$ meson. An example is the hard spectator term 
in the QCD factorization formula for the exclusive decay $B\to K^*\gamma$ 
\cite{Beneke:2001at,Bosch:2001gv}. The relevant SCET operators can be 
taken as \cite{Hill:2002vw}
\begin{eqnarray}\label{QRdef}
   Q_{(C)}(s,t)
   &=& \big[ \bar\xi\,W_c \big](s\bar n)\,\frac{\nbslash}{2}\,
    \Gamma_1\,T_1\,\big[ W_c^\dagger\,\xi \big](0)\,\,
    \big[ \bar q_s\,S_s \big](tn)\,\frac{\nslash}{2}\,\Gamma_2\,
    T_2\,\big[ S_s^\dagger\,h \big](0) \nonumber\\
   &\equiv& \int_0^\infty\!d\omega\,e^{-i\omega t}
    \int_0^{\,\bar n\cdot P}\!d\sigma\,e^{i\sigma s}\,
    Q_{(C)}(\omega,\sigma) \,,
\end{eqnarray}
where the color label $C=S$ or $O$ refers to the color singlet-singlet 
and color octet-octet structures $T_1\otimes T_2={\bf 1}\otimes{\bf 1}$ 
or $T_A\otimes T_A$, respectively. The quantity $\bar n\cdot P$ is the 
total momentum carried by all collinear particles, which is fixed by
kinematics. (Strictly speaking, this is a momentum operator.) The 
matrices $\Gamma_i$ represent any of the Dirac basis matrices. Between 
two collinear fields only the three possibilities 
$\Gamma_1=1,\gamma_5,\gamma_\perp^\mu$ are allowed, whereas $\Gamma_2$ is 
not constrained. The factor $\nslash/2$ between the fields $\bar q_s$ and 
$h$ ensures that the $B$-meson matrix element of the soft-quark current 
can be expressed in terms of the leading-order $B$-meson LCDA 
$\phi_+^B(\omega,\mu)$ \cite{Grozin:1996pq}. In light-cone gauge, 
$\omega=n\cdot p_s$ corresponds to the plus component of the momentum of 
the spectator anti-quark in the $B$ meson, while 
$\sigma=\bar n\cdot p_\xi$ denotes the minus component of the momentum 
of the quark inside a light final-state meson. It is conventional to 
introduce a dimensionless variable $u=\sigma/\bar n\cdot P\in[0,1]$ 
corresponding to the longitudinal momentum fraction carried by the quark.

\begin{figure}[t]
\begin{center}
\begin{tabular}{ccc}
\raisebox{\gw}{\includegraphics[width=\gw,angle=270]{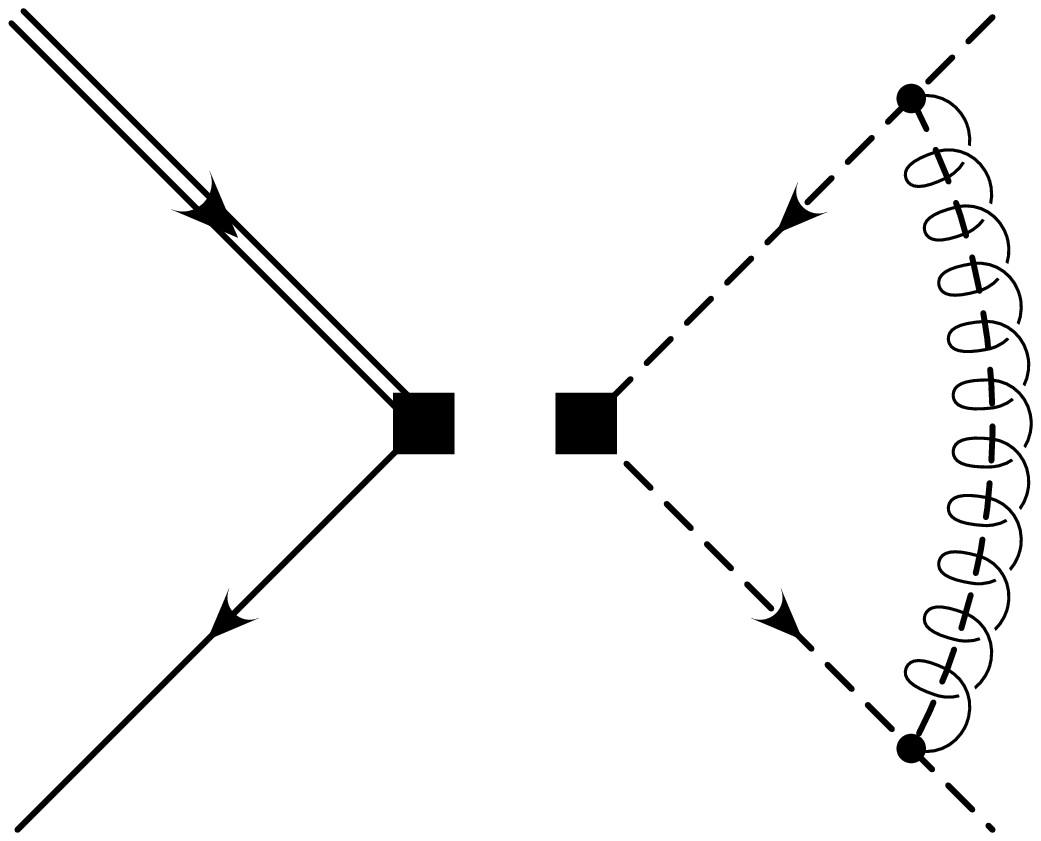}} &
\raisebox{\gw}{\includegraphics[width=\gw,angle=270]{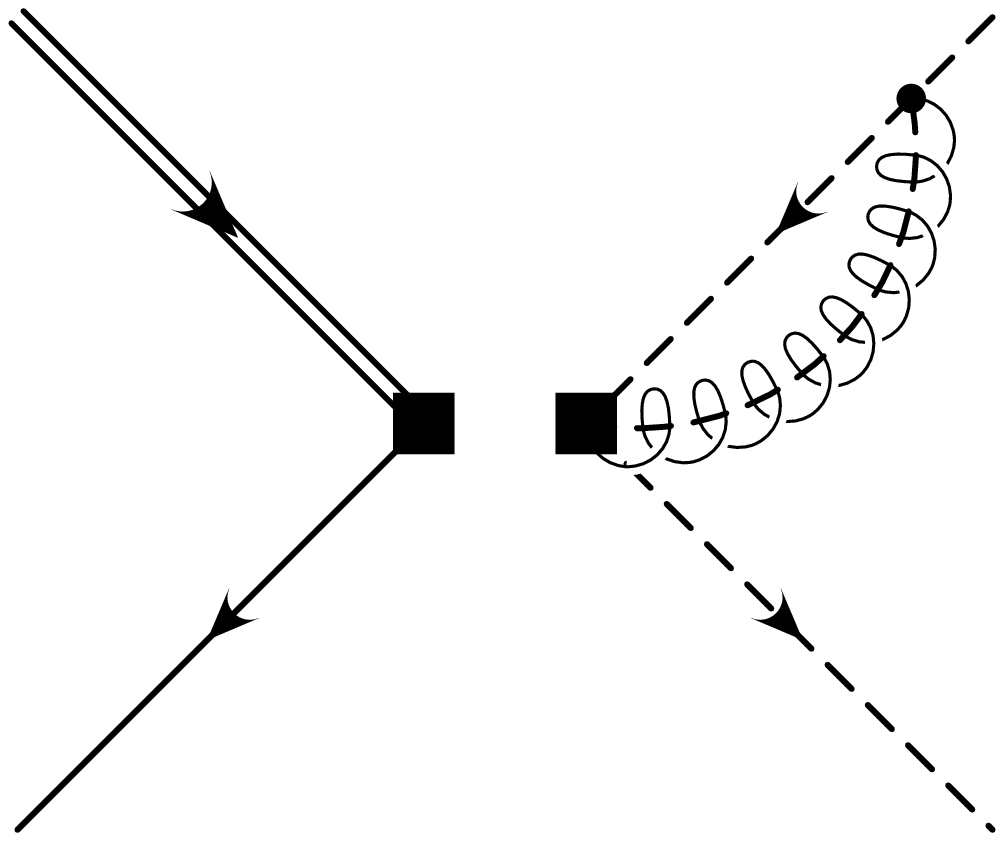}} &
\raisebox{\gw}{\includegraphics[width=\gw,angle=270]{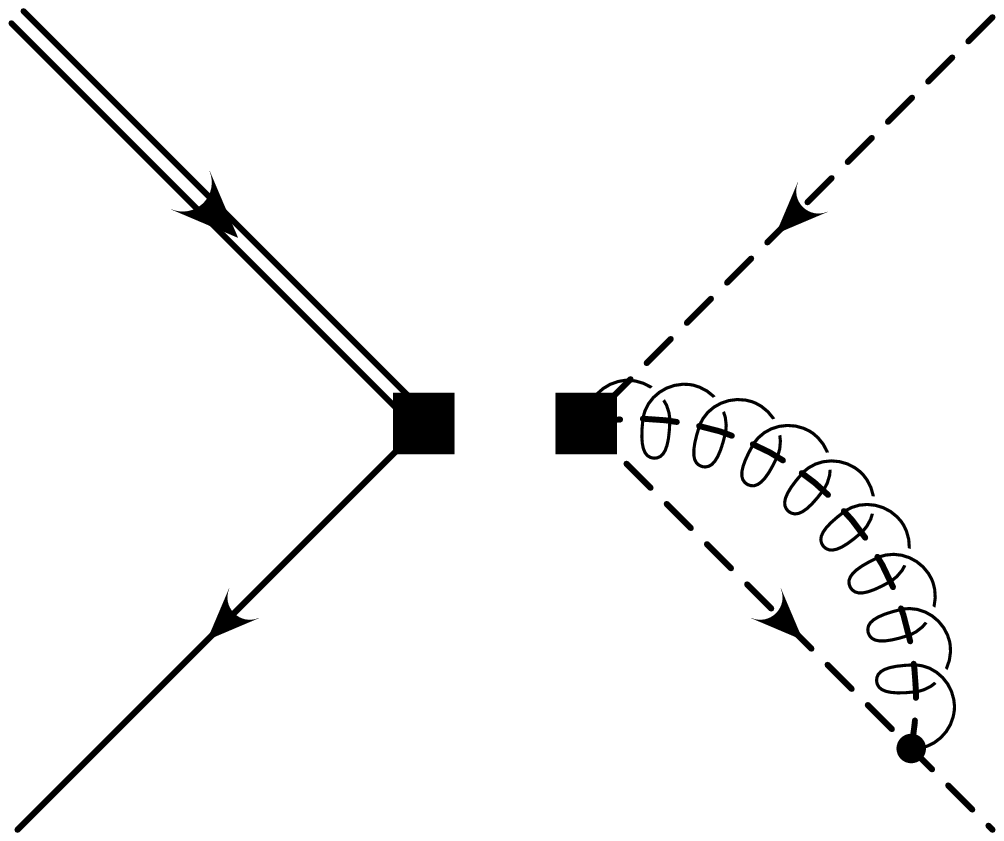}} \\ && \\
\raisebox{\gw}{\includegraphics[width=\gw,angle=270]{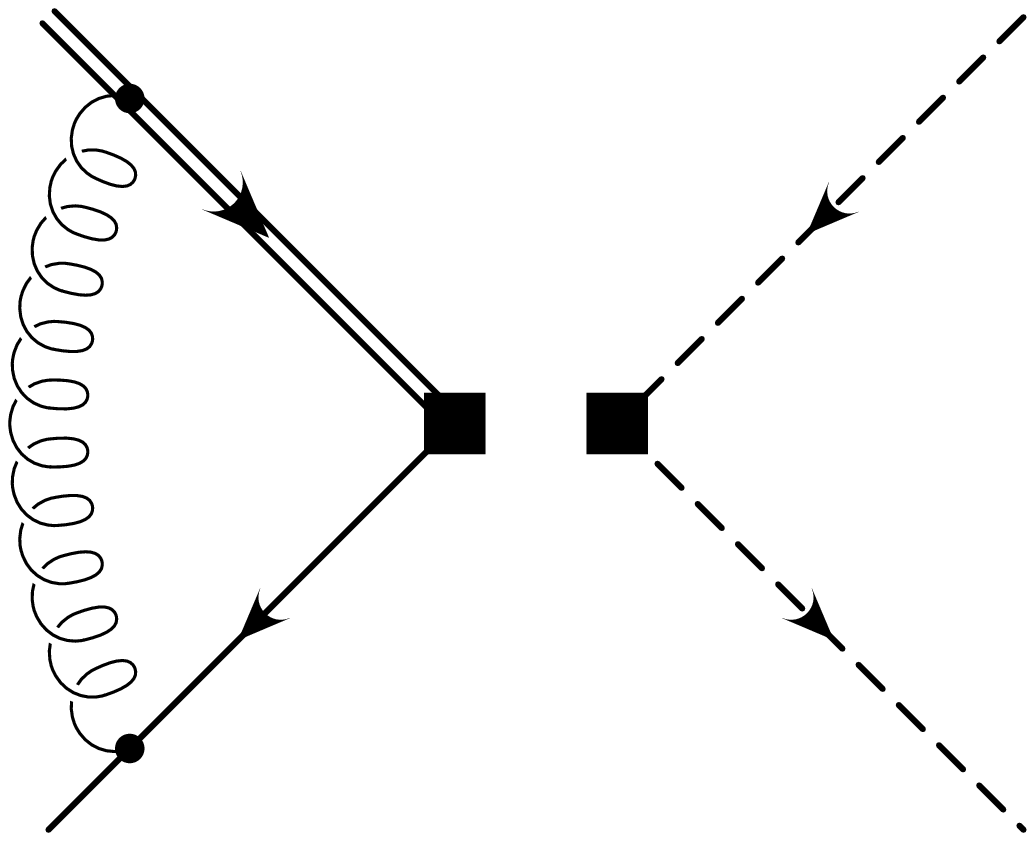}} &
\raisebox{\gw}{\includegraphics[width=\gw,angle=270]{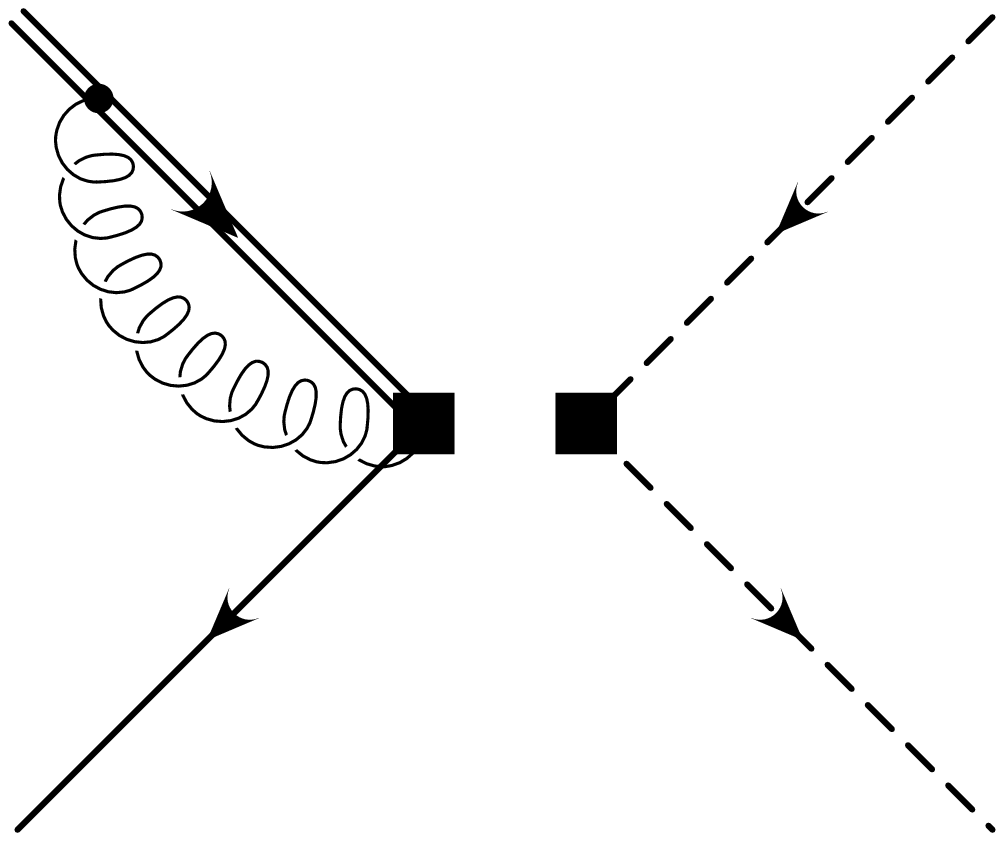}} &
\raisebox{\gw}{\includegraphics[width=\gw,angle=270]{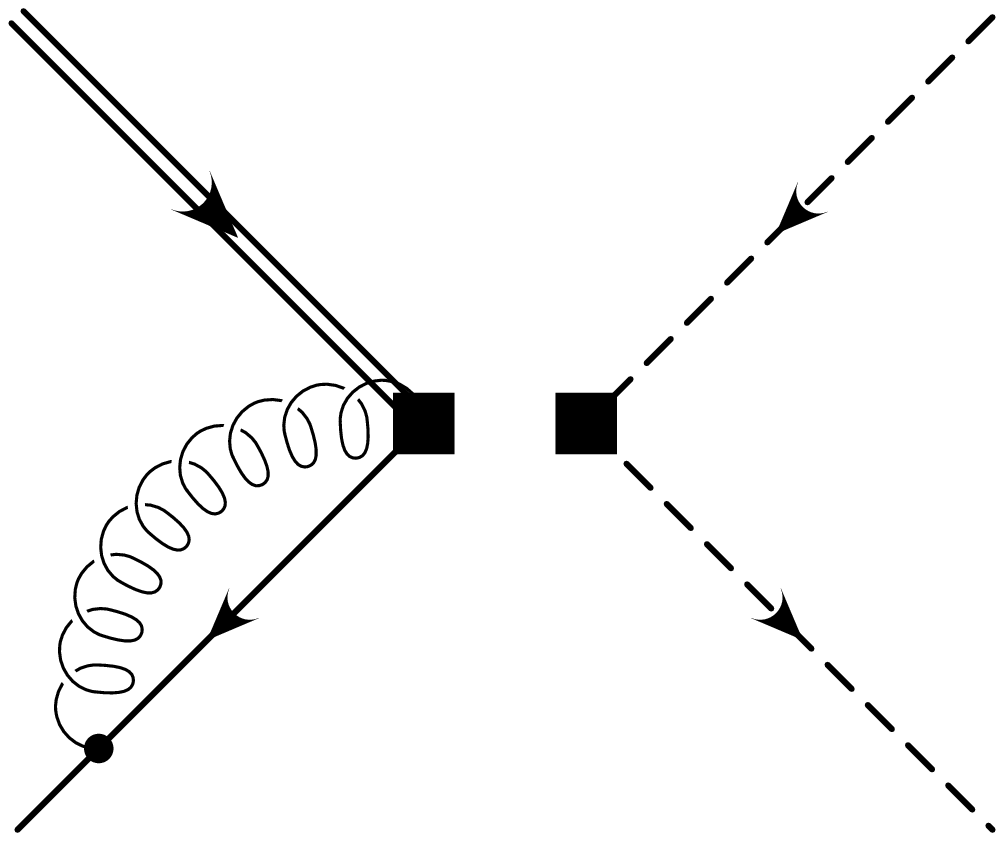}}\\  
&&
\end{tabular}
\begin{tabular}{cccc}
\raisebox{\gw}{\includegraphics[width=\gw,angle=270]{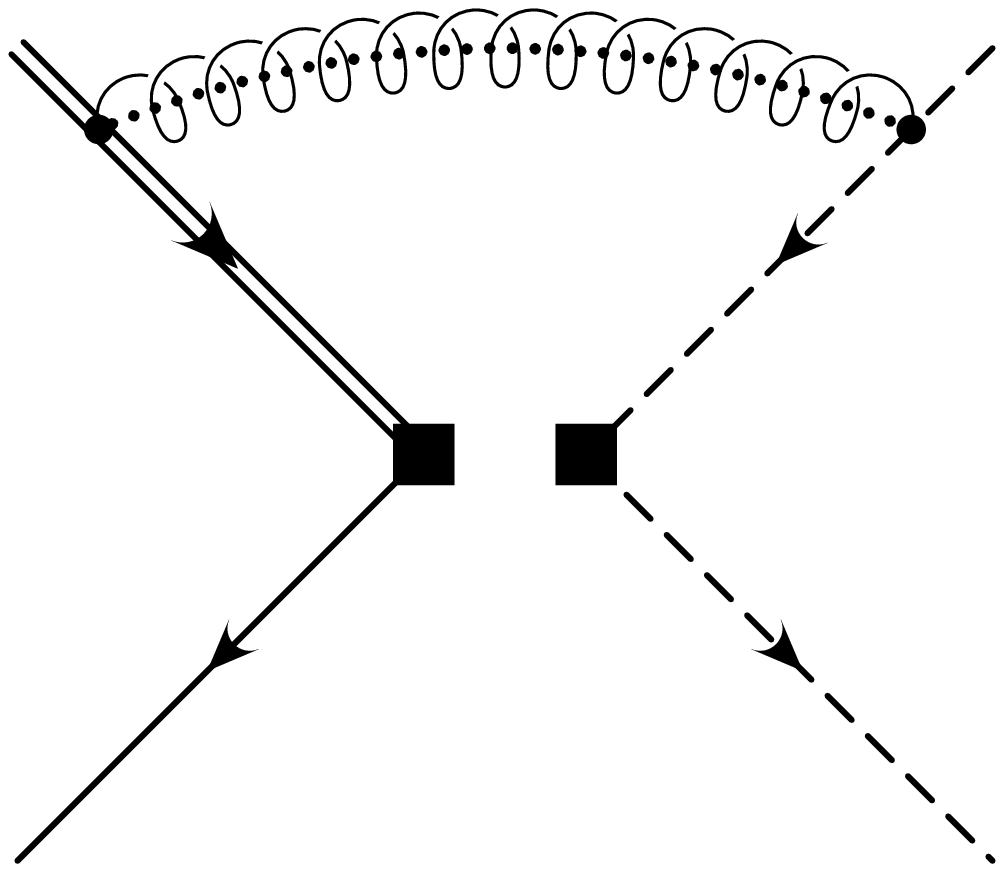}} &
\raisebox{\gw}{\includegraphics[width=\gw,angle=270]{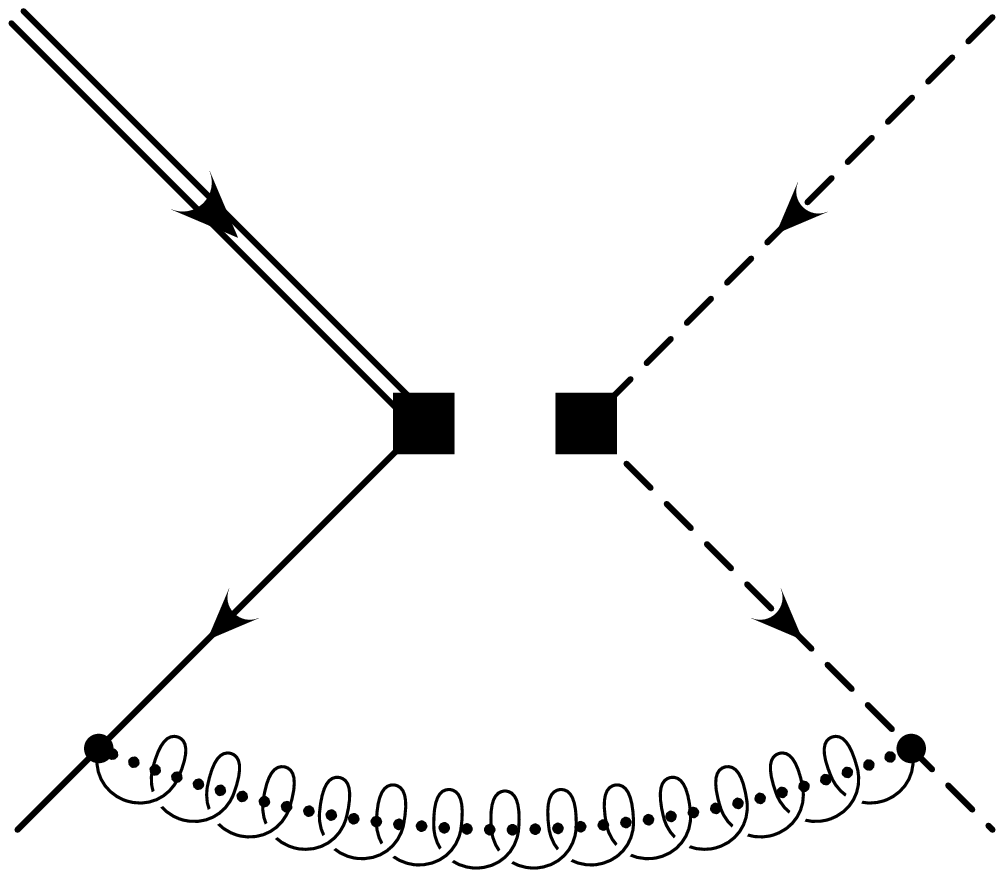}} &
\raisebox{\gw}{\includegraphics[width=\gw,angle=270]{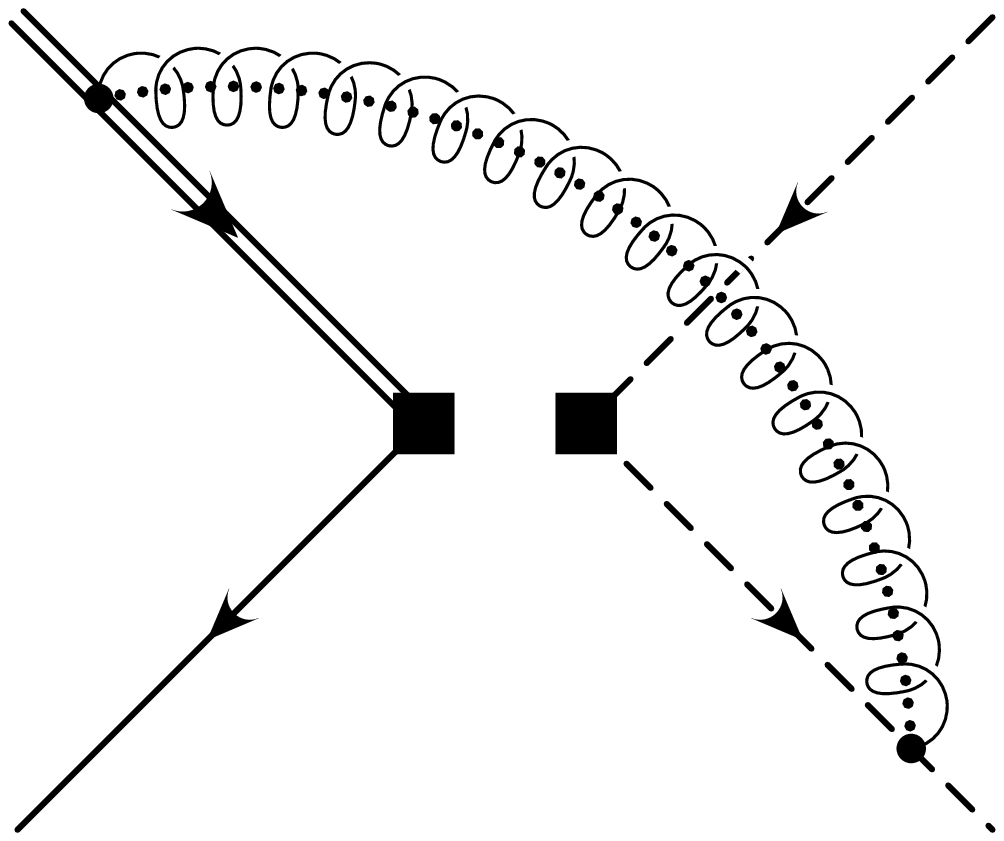}} &
\raisebox{\gw}{\includegraphics[width=\gw,angle=270]{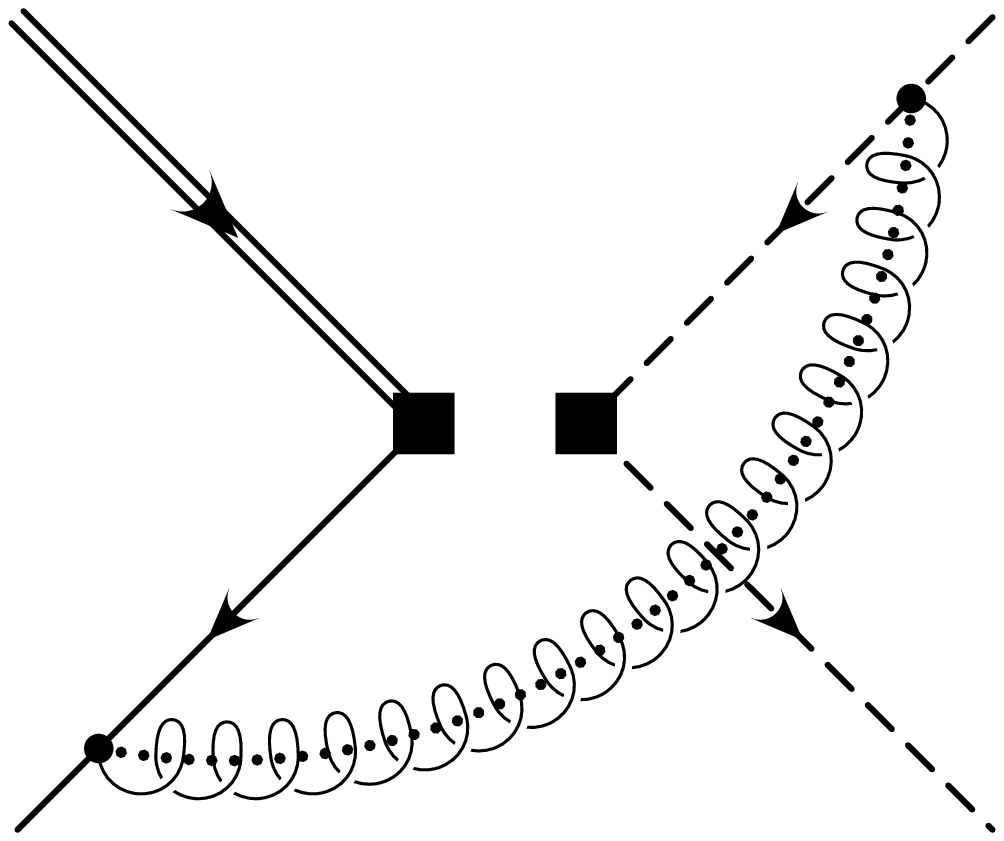}}\\
\end{tabular}
\end{center}
\vspace{-0.2cm}
\centerline{\parbox{14cm}{\caption{\label{fig:gamma}
SCET graphs contributing to the anomalous dimension of the four-quark 
operators $Q_{(C)}(\omega,\sigma)$. Full lines denote soft fields, dashed 
lines collinear fields, and dotted lines soft-collinear fields.}}}
\end{figure}

The momentum-space operators $Q_{(C)}(\omega,\sigma)$ obey the 
integro-differential RG equation
\begin{equation}
   \frac{d}{d\ln\mu}\,Q_{(C)}(\omega,u\,\bar n\cdot P)
   = - \int_0^{\infty}\!d\omega' \int_0^1\!du'\,
   \gamma_{(C)}(\omega,\omega',u,u',\bar n\cdot P,\mu)\,
   Q_{(C)}(\omega',u'\,\bar n\cdot P) \,.
\end{equation}
To obtain the anomalous dimensions at leading order we compute the 
$1/\epsilon$ poles of the diagrams shown in Figure~\ref{fig:gamma} in
dimensional regularization and add the contributions from wave-function 
renormalization. Note that only soft-collinear gluons can be exchanged 
between the soft and collinear currents. For the color-singlet case 
$T_1\otimes T_2={\bf 1}\otimes{\bf 1}$ we find that the sum of the four 
diagrams with soft-collinear exchanges (but not each diagram separately) 
is  UV finite. The anomalous dimension is then a combination of the 
anomalous dimensions for the two non-local currents in (\ref{QRdef}). At 
one-loop order we obtain
\begin{equation}
   \gamma_{(S)}(\omega,\omega',u,u',\bar n\cdot P,\mu)
   = \frac{C_F\alpha_s}{\pi}\,\Big[ \delta(\omega-\omega')\,V(u,u') 
   + \delta(u-u')\,H(\omega,\omega',\mu) \Big] \,,
\end{equation}
where (with $\bar u\equiv 1-u$) 
\begin{eqnarray}
   V(u,u') 
   &=& - \left[
    \frac{u}{u'} \left( \frac{1}{u'-u} + c(\Gamma_1) \right) \theta(u'-u) 
    + \frac{\bar u}{\bar u'} \left( \frac{1}{u-u'} + c(\Gamma_1) \right) 
    \theta(u-u') \right]_+ \nonumber\\
   &&\mbox{}\!+ \frac{1-c(\Gamma_1)}{2}\,\delta(u-u')
\end{eqnarray}
with $c(1)=c(\gamma_5)=1$, $c(\gamma_\perp^\mu)=0$ is the Brodsky--Lepage
kernel \cite{Lepage:1980fj} for the evolution of the leading-twist LCDA 
of a light meson, which we have reproduced here using the Feynman rules 
of SCET. The plus distribution is defined as
\begin{equation}
   [f(u,u')]_+ = f(u,u') - \delta(u-u') \int_0^1\!dw\,f(w,u') \,,
\end{equation}
which coincides with the conventional definition if the distribution acts 
on functions $g(u)$ but not if it acts on functions $g(u')$. The function
\begin{equation}
   H(\omega,\omega',\mu)
   = \left( \ln\frac{\mu\,v\cdot n}{\omega} - \frac54 \right)
   \delta(\omega-\omega')
   - \omega \left[ \frac{\theta(\omega-\omega')}{\omega(\omega-\omega')}
   +  \frac{\theta(\omega'-\omega)}{\omega'(\omega'-\omega)} \right]_+
\end{equation}
is the analogous kernel governing the evolution of the leading-order
$B$-meson LCDA \cite{Lange:2003ff}. Here the plus distribution is 
symmetric in the two arguments and defined as
\begin{equation}
   \int_0^\infty\!d\omega'\,[f(\omega,\omega')]_+\,g(\omega')
   =  \int_0^\infty\!d\omega'\,f(\omega,\omega')\,
   \big[ g(\omega') - g(\omega) \big] \,.
\end{equation}

For the color-octet case $T_1\otimes T_2=T_A\otimes T_A$ things are more
complicated. In this case the diagrams in the first two lines of 
Figure~\ref{fig:gamma} reproduce the singlet anomalous dimension except 
for a different overall color factor, but in addition these graphs contain 
$1/\epsilon$ poles that depend on the IR regulators. In units of the 
tree-level matrix element $\langle\,Q_{(O)}(\omega,\sigma)\,\rangle$, the
extra terms are
\begin{equation}\label{term1}
   \frac{N}{2}\,\frac{\alpha_s}{2\pi} \left\{
   \left( \frac{3}{2\epsilon^2}
   - \frac{1}{\epsilon}\,\ln\frac{-2v\cdot l_h}{\mu}
   - \frac{1}{\epsilon}\,\ln\frac{-l_q^2}{\mu}
   + \frac{5}{4\epsilon} \right)
   + \bigg( \frac{2}{\epsilon^2}
   - \frac{1}{\epsilon}\,\ln\frac{-p_\xi^2}{\mu^2}
   - \frac{1}{\epsilon}\,\ln\frac{-p_{\bar\xi}^2}{\mu^2}
   + \frac{3}{2\epsilon} \bigg) \right\} ,
\end{equation}
where $l_i$ are the incoming soft momenta, and $p_i$ denote the outgoing 
collinear momenta. The first parenthesis shows the soft contribution,
while the second one gives the collinear contribution. In addition, in 
the color-octet case the sum of the soft-collinear exchange graphs shown
in the last line in Figure~\ref{fig:gamma} does not vanish, but adds up to
\begin{equation}\label{term2}
   - \frac{N}{2}\,\frac{\alpha_s}{2\pi} \left(
   \frac{2}{\epsilon^2}
   - \frac{1}{\epsilon}\,
   \ln\frac{(-2v\cdot l_h)(-l_q^2)(-p_\xi^2)(-p_{\bar\xi}^2)}
           {(n\cdot v)(n\cdot l_q)(\bar n\cdot p_\xi)
            (\bar n\cdot p_{\bar\xi})\,\mu^4} \right) .
\end{equation}
In the sum of the two terms (\ref{term1}) and (\ref{term2}) the 
dependence on the IR regulators drops out. Our final result for the 
anomalous dimension in the octet case is
\begin{eqnarray}
   \gamma_{(O)}(\omega,\omega',u,u',\bar n\cdot P,\mu)
   &=& - \frac{1}{2N}\,\frac{\alpha_s}{\pi}\,\Big[
    \delta(\omega-\omega')\,V(u,u') 
    + \delta(u-u')\,H(\omega,\omega',\mu) \Big] \\
   &&\mbox{}\!- \frac{N}{2}\,\frac{\alpha_s}{\pi}\,
    \delta(\omega-\omega')\,\delta(u-u') \left(
    \ln\frac{\mu^3}{n\cdot v\,\omega\,(\bar n\cdot P)^2}
    - \ln u\bar u + \frac{11}{4} \right) . \nonumber
\end{eqnarray}

\section{Decoupling transformation\label{sec:decoupling}}

The leading-order interactions between soft-collinear fields and soft or 
collinear fields in the SCET Lagrangian (\ref{Lscet}) can be removed by 
a redefinition of the soft and collinear fields \cite{Becher:2003qh}. In 
analogy with the decoupling of ultra-soft gluons in SCET$_{\rm I}$ 
\cite{Bauer:2001yt}, we define new fields
\begin{equation}\label{another}
\begin{aligned}
   q_s(x) = W_{sc}(x_+)\,q_s^{(0)}(x) \,, \qquad
   h(x) = W_{sc}(x_+)\,h^{(0)}(x) \,, \qquad
   \xi(x) = S_{sc}(x_-)\,\xi^{(0)}(x) \,, \\
   A_s^\mu(x) = W_{sc}(x_+)\,A_s^{(0)\mu}(x)\,W_{sc}^\dagger(x_+)
    \,, \qquad 
   A_c^\mu(x) = S_{sc}(x_-)\,A_c^{(0)\mu}(x)\,S_{sc}^\dagger(x_-) \,.
    \quad~
\end{aligned}
\end{equation}
The quantities $W_{sc}$ and $S_{sc}$ are yet another set of Wilson lines. 
They are defined in analogy with $W_c$ and $S_s$ in (\ref{WSdef}), 
however with the gluon fields replaced by soft-collinear gluon fields in 
both cases. These objects are invariant under soft and collinear gauge 
transformations, while under a soft-collinear gauge transformation they 
transform as
\begin{equation}
   W_{sc}(x_+)\to U_{sc}(x_+)\,W_{sc}(x_+) \,, \qquad
   S_{sc}(x_-)\to U_{sc}(x_-)\,S_{sc}(x_-) \,.
\end{equation}
Consequently, the new fields with ``(0)'' superscripts are invariant 
under soft-collinear gauge transformations. When they are introduced in 
the SCET Lagrangian the terms ${\cal L}_s$, ${\cal L}_c$, and 
${\cal L}_{sc}$ retain their original form, while the leading-order 
interaction Lagrangian ${\cal L}_{\rm int}^{(0)}$ vanishes. Residual 
interactions between soft-collinear and soft or collinear fields start at 
$O(\lambda)$ \cite{Becher:2003qh}. After the field redefinition it is 
convenient to introduce the gauge-invariant building blocks 
\cite{Hill:2002vw}
\begin{equation}
\begin{aligned}
   \Q_s(x) &= S_s^{(0)\dagger}(x)\,q_s^{(0)}(x)
    = W_{sc}^\dagger(x_+)\,S_s^\dagger(x)\,q_s(x) \,, \\
   \H(x) &= S_s^{(0)\dagger}(x)\,h^{(0)}(x)
    = W_{sc}^\dagger(x_+)\,S_s^\dagger(x)\,h(x) \,, \\
   \X(x) &= W_c^{(0)\dagger}(x)\,\xi^{(0)}(x)
    = S_{sc}^\dagger(x_-)\,W_c^\dagger(x)\,\xi(x) \,,
\end{aligned}
\end{equation}
which are invariant under all three types of gauge transformations. 

The fact that interactions of soft-collinear fields with other fields
can be decoupled from the strong-interaction Lagrangian does not
necessarily imply that these fields can be ignored at leading order in
power counting. The question is whether the decoupling transformation
(\ref{another}) leaves external operators such as weak-interaction
currents invariant. The analysis of the previous sections indicates
that in some cases the soft-collinear exchange graphs contribute to
the calculation of the anomalous dimensions. Let us then study what
happens when the decoupling transformation is applied to the various
types of operators.

Under the transformation (\ref{another}), the soft-collinear currents in 
(\ref{hc}) and (\ref{sc}) transform into (setting $x_\perp=0$ for 
simplicity)
\begin{equation}
\begin{aligned}
   \big[ \bar\xi\,W_c \big](x_+)\,\Gamma\,\big[ S_s^\dagger\,h \big](x_-)
   &\to \bar\X(x_+)\,S_{sc}^\dagger(0)\,\Gamma\,W_{sc}(0)\,\H(x_-) \,, \\
   \big[ \bar\xi\,W_c \big](x_+)\,\Gamma\,
   \big[ S_s^\dagger\,q_s\big](x_-)
   &\to \bar\X(x_+)\,S_{sc}^\dagger(0)\,\Gamma\,W_{sc}(0)\,\Q_s(x_-) \,.
\end{aligned}
\end{equation}
We observe that the soft-collinear fields do not decouple from these 
currents but rather form a light-like Wilson loop with a cusp at $x=0$. 
The anomalous dimension of the combination $S_{sc}^\dagger\,W_{sc}$ is 
the universal cusp anomalous dimension times a logarithm of the 
soft-collinear scale, see the last terms in the first lines in 
(\ref{eq:hc}) and (\ref{eq:sc}). After adding the contributions from the 
soft and collinear sectors, the dependence on the IR regulators drops 
out. However, the coefficient of the logarithm of $v\cdot p_{c-}$ in the 
heavy-collinear current and $p_{s+}\!\cdot p_{c-}$ in the soft-collinear 
current is unchanged, since both the soft and the collinear part are 
independent of these large scales. This cancellation also explains why 
the anomalous dimensions of the soft-collinear and heavy-collinear 
currents involve $-\Gamma_{\rm cusp}$ and $-\frac12\Gamma_{\rm cusp}$, 
respectively:
\begin{equation}
\begin{aligned}
   \Gamma_{\rm cusp} \left[
   \ln\frac{2p_{s+}\!\cdot p_{c-}\,\mu^2}{(-p_s^2)(-p_c^2)}
   + \ln\frac{-p_s^2}{\mu^2} + \ln\frac{-p_c^2}{\mu^2} \right]
   &= - \Gamma_{\rm cusp}\,\ln\frac{\mu^2}{2p_{s+}\!\cdot p_{c-}}
    \,, \\
   \Gamma_{\rm cusp} \left[
   \ln\frac{2v\cdot p_{c-}\,\mu^2}{(-2v\cdot p_s)\,(-p_c^2)}
   + \ln\frac{-2v\cdot p_s}{\mu} + \ln\frac{-p_c^2}{\mu^2} \right]
   &= -\frac12\,\Gamma_{\rm cusp}\,\ln\frac{\mu^2}{(2v\cdot p_{c-})^2}
    \,.
\end{aligned}
\end{equation}
Similar arguments were used by Korchemsky in his analysis of the 
off-shell Sudakov form factor \cite{Korchemsky:1988hd}.

The effect of the field redefinition (\ref{another}) on the four-quark
operators is different. The color singlet-singlet operator is invariant, 
namely (setting $x=0$ for simplicity)
\begin{equation}\label{singlet}
   Q_{(S)}(s,t)\to \bar\X(s\bar n)\,\Gamma_1\,\X(0)\,\,
   \bar\Q_s(tn)\,\Gamma_2\,\H(0) \,,
\end{equation}
since the additional soft-collinear Wilson lines come in pairs 
$W_{sc}^\dagger\,W_{sc}=1$ and $S_{sc}^\dagger\,S_{sc}=1$. The color 
octet-octet operator is however not invariant. It transforms into
\begin{equation}
   Q_{(O)}(s,t)\to \bar\X(s\bar n)\,\Gamma_1\,
   \big[ S_{sc}^\dagger\,T_A\,S_{sc} \big](0)\,\X(0)\,\,
   \bar\Q_s(tn)\,\Gamma_2\,
   \big[ W_{sc}^\dagger\,T_A\,W_{sc} \big](0)\,\H(0) \,.
\end{equation}
Because the objects $W_{sc}^\dagger\,T_A\,W_{sc}$ and 
$S_{sc}^\dagger\,T_A\,S_{sc}$ are pure color octets, the result can be 
rewritten as
\begin{equation}\label{octet}
   Q_{(O)}(s,t)\to c_{AB}[A_{sc}]\,\bar\X(s\bar n)\,\Gamma_1\,T_A\,\X(0)\,
   \,\bar\Q_s(tn)\,\Gamma_2\,T_B\,\H(0) \,,
\end{equation}
where
\begin{equation}
  c_{AB}[A_{sc}]
  = 2\,\mbox{Tr}\big[ S_{sc}\,T_A\,S_{sc}^\dagger\,
  W_{sc}\,T_B\,W_{sc}^\dagger \big](0)
\end{equation}
is a functional of the soft-collinear gluon field. Since after the 
decoupling transformation the SCET Lagrangian no longer contains 
leading-order interactions between soft-collinear and soft or collinear 
fields, it follows that at leading power the soft, collinear, and 
soft-collinear parts of the current operators in (\ref{singlet}) and 
(\ref{octet}) only interact among themselves. The color structure of the 
color-singlet and color-octet currents built up of soft or collinear 
fields are preserved in these interactions. Hence, both types of 
four-quark operators are multiplicatively (in the convolution sense) 
renormalized in the effective theory -- unlike in full QCD, where they 
mix under renormalization.

The presence of the functional $c_{AB}[A_{sc}]$ in the octet case 
explains why soft-collinear modes give a non-zero contribution to the 
anomalous dimension of the operator $Q_{(O)}$. However, since this 
operator does not mix into the singlet-singlet operator $Q_{(S)}$, this 
effect does not propagate into physical decay amplitudes (as hadronic 
matrix elements of color-octet currents vanish). The decoupling of 
soft-collinear fields from the color singlet-singlet operator implies 
that, to all orders in perturbation theory, the anomalous dimension of 
the four-quark operator $Q_{(S)}$ is the sum of the anomalous dimensions 
of the two currents $\bar\X(s\bar n)\,\Gamma_1\,\X(0)$ and 
$\bar\Q_s(tn)\,\Gamma_2\,\H(0)$, in accordance with the one-loop result 
obtained in the previous section. This observation has important 
implications for applications of SCET to proofs of QCD factorization 
theorems. For instance, the potential effects of soft-collinear modes 
have been ignored in studies of factorization for the exclusive decay 
$B\to D\pi$ \cite{Beneke:2000ry,Bauer:2001cu}. Our results justify this 
treatment {\em a posteriori}, thereby completing the proof of 
factorization for this decay.

\section{Conclusions}

We have argued that soft-collinear effective theory (SCET) for processes 
involving both soft and collinear partons is a more complicated (yet more
interesting) theory than previously assumed. In addition to soft and 
collinear particles, which make up the external hadron states, there 
exist soft-collinear messenger modes, which can communicate between the 
soft and collinear sectors of the theory. The presence of these modes, 
and the fact that they have leading-order interactions with both soft and 
collinear particles, destroys the trivial factorization of soft and 
collinear physics that was thought to be a property of the effective 
theory. As a consequence, a careful analysis of soft-collinear exchange
contributions must be part of any proof of QCD factorization theorems.

We have extended the construction of the SCET Lagrangian to include
external sources such as current and four-quark operators containing
both soft and collinear fields. To build up confidence in the
soft-collinear modes, we have explicitly shown that they are needed
for obtaining the correct ultraviolet behavior of effective theory
amplitudes. The explicit examples we have investigated show that only
the sum of soft, collinear and soft-collinear contributions to an
amplitude is physically meaningful. In cases where the soft-collinear
modes cannot be decoupled by field redefinitions, SCET amplitudes
become sensitive to the large scale $E$ through the particular scaling
$m_{sc}^2\sim\Lambda^3/E$ of soft-collinear momenta.  Only part of
this sensitivity is of a short-distance nature, as described by the
anomalous dimensions of SCET operators. In addition, amplitudes may
contain a long-distance dependence on the large scale that cannot be
factorized using renormalization-group techniques.

In the strong-interaction sector of SCET the leading-order interactions 
of soft-collinear fields with soft or collinear fields can be removed 
using field redefinitions, leaving residual interactions that are power
suppressed. In the presence of external operators this decoupling 
property is no longer guaranteed, but depends on whether external 
operators remain invariant under the decoupling transformation. We have 
shown, for instance, that currents containing a soft and a collinear 
quark are not invariant, implying that the effects of 
soft-collinear contributions to current matrix elements must be studied
carefully.

In summary, we have completed the discussion of the SCET Lagrangian at 
leading power and including external operators relevant to weak 
interactions. The framework developed here forms the basis for 
systematic, complete proofs of QCD factorization theorems for exclusive 
$B$-meson decay amplitudes. In particular, our finding of the decoupling 
of soft-collinear contributions for color singlet-singlet four-quark 
operators completes the proof of factorization for the decay $B\to D\pi$.

\vspace{0.5cm}\noindent
{\em Acknowledgments:\/}
The work of T.B.\ and R.J.H.\ is supported by the Department of Energy 
under Grant DE-AC03-76SF00515. The research of B.O.L.\ and M.N.\ is 
supported by the National Science Foundation under Grant PHY-0098631.

\end{document}